\documentclass{kluwer}    

\begin{document}                                                                                   
\begin{article}
\begin{opening}    
\title{Theoretical Models for Producing Circularly Polarized Radiation in Extragalactic Radio Sources}
      
\author{John F. C. \surname{Wardle}}  
\institute{Brandeis University\\Waltham, MA 02454, U.S.A.}
\author{Daniel C. \surname{Homan}}  
\institute{National Radio Astronomy Observatory\thanks{The National Radio Astronomy Observatory is a facility of the National Science Foundation operated under cooperative agreement by Associated Universities, Inc.}\\Charlottesville, VA 22903, U.S.A.}
\runningauthor{Wardle and Homan}
\runningtitle{Theoretical Models for Producing Circularly Polarized Radiation}

\date{December 18, 2002}

\begin{abstract}
We discuss the production of circular polarization in compact radio sources both by the intrinsic mechanism and by Faraday conversion. We pay particular attention to the magnetic field structure, considering partially ordered fields and Laing sheets, and distinguishing between uniform and unidirectional fields. (The latter can be constrained by flux conservation arguments.) In most cases, Faraday conversion is the more important mechanism. Conversion operates on Stokes U, which can be generated by internal Faraday rotation, or by magnetic field fluctuations, which can therefore produce circular polarization even in a pure pair plasma. We also show that the spectrum of circular polarization in an inhomogeneous jet can be quite different from that in a uniform source, being flat or even inverted.
 
\end{abstract}

\end{opening}

\section{Introduction}
The resurgence of interest in circular polarization (CP), a result of which is this workshop, stems partly from observational advances. These include the impressive precision of the ATCA (Rayner, Norris \& Sault 2000), and the ability to determine the location of the CP with milliarcsecond resolution (e.g. Homan \& Wardle 1999). CP observations are also being made with the VLA (Bower, Falke \& Backer 1999), Westerbork (e.g. de Bruyn, these proceedings), and at the University of Michigan (Aller \& Aller, these proceedings).

The theoretical interpretation of these observations is unfamiliar to many people, partly because of the omission of CP from standard texts which cover synchrotron radiation. Jones and O'Dell (1977) write down an analytic solution to the coupled equations of transfer for all four Stokes parameters in a uniform source. Jones (1988) carried out numerical simulations of an inhomogeneous jet containing a completely tangled field. He found that CP was invariably produced by Faraday conversion, at surprisingly high local fractional levels. His paper stimulated us to develop techniques to detect CP at VLBI resolution.

Here we discuss the production of CP in compact radio sources, paying particular attention to the magnetic field structure. We restrict ourselves to the intrinsic mechanism, and to Faraday conversion. The aim is to develop expressions that may be applied to real sources. For simplicity, we only discuss the optically thin, Faraday thin cases. Jones \& O'Dell (1977) and Ruszkowski \& Begelman (2002) derive more general expressions appropriate to high optical and Faraday depth.

In the final section we discuss the effect of inhomogeneous jet structure on the spectrum of CP, and show that it can be large. In some sources, CP might be strongest at short millimeter wavelengths, where we are observing as close the the base of the jet as possible,

\section{Magnetic Field Structures}

In order to calculate the expected polarization of a synchrotron emitting region, it is essential to consider carefully the structure of the magnetic field. Since the observed linear polarization in most parts of most radio sources is far less than the $ 70\%$ or so predicted for optically thin synchrotron radiation in a uniform magnetic field, it is clear that the magnetic field is highly tangled, with a correlation length much smaller than a beamwidth. For circular polarization, it is also important to consider the directionality of the magnetic field. A perfectly ordered ``uniform'' field may also be unidirectional, or it may consist of sheared loops and have an equal number of field lines going in either direction. In both cases, the fractional linear polarization will be $\sim 70\%$, but the circular polarization will be a maximum in the first, but zero in the second.

A very useful family of magnetic field structures is due to originally Laing (1980), in which a completely tangled magnetic field is compressed into a plane (often called a Laing sheet). Hughes, Aller \& Aller (1985) generalized this for an arbitrary amount of compression to represent shocks of various strengths. Wardle et al. (1994) added a uniform component of magnetic field to represent the longitudinal magnetic field often seen in high luminosity jets.

In these models, we begin with a randomly oriented magnetic field of amplitude $B_r$, together with a uniform magnetic field $B_u$ aligned with the jet. The magnetic field is then compressed transverse to the jet direction, so that unit length is compressed to length $k$. The line of sight to the observer, in the rest frame of the emitting fluid, makes an angle $\epsilon$ to the plane of compression (or $90 - \epsilon$ to the jet direction). The electron energy distribution is $N(\gamma) = K\gamma^{-2\alpha - 1}$, for $\gamma_i < \gamma < \gamma_j$, where $K$, $\gamma_i$ and $\gamma_j$ may all depend on the compression factor $k$. The spectral index $\alpha$ is defined so that $S \propto \nu^{-\alpha}$.

The resulting Stokes parameters are (Wardle et al. 1994):
\begin{eqnarray}
\langle I \rangle & \propto & K(k) k^{-2} B_{r}^{2}\{2+[3k^{2} \xi^{2}-(1-k^{2})]\cos^{2}\epsilon\} \\
\langle Q \rangle & \propto & m_{0} K(k) k^{-2} B_{r}^{2} [3k^{2} \xi^{2}-(1-k^{2})]\cos^{2}\epsilon \\
\langle U \rangle  & \propto & 0 
\end{eqnarray}
where $m_0 = (3\alpha +3)/(3\alpha + 5)$, and $\xi = B_u / B_r$.
The coordinate system is chosen so that electric vectors transverse to the jet are positive Stokes Q. To make the integrals tractable, we have assumed that $\alpha = 1$ in the angle averaging, but kept the proper spectral index dependencies elsewhere. The error in this customary approximation is evaluated in Wardle et al (1994).

The mean value of Stokes U is zero by symmetry, because it is equally likely to be positive or negative. The r.m.s. value of U (i.e. the width of the distribution) is not zero, and this is a useful quantity to calculate:
\begin{equation}
U_{rms}  \propto  m_{0} K(k) k^{-2} B_{r}^{2} \frac{2}{\sqrt{15}} [\sin^{2}\epsilon + k^2\cos^{2}\epsilon + 5k^2 \xi^2 \cos^{2}\epsilon]^{\frac{1}{2}} 
\end{equation}

This will be the typical magnitude of the U from a single coherence length or cell of the field structure described above. If there are N such regions within a resolution element, then the emerging U is reduced by $\sqrt{N}$, and the fluctuation in electric vector position angle (EVPA) is $\Delta\chi \sim U_{rms}/(2\sqrt{N}\langle Q \rangle)$.

Of course, the magnetic field structure described here is only one of many possibilities. But it is reasonably easy to compute analytic results for it, and it is flexible enough to represent a variety of plausible configurations. The addition of the uniform component, $B_u$, is not necessary to produce a net magnetic field aligned with the jet, since one may stretch the field by setting $k > 1$, which accomplishes much the same thing. The main reason we include $B_u$ is to allow for a component of the magnetic field that may also be {\em unidirectional}. This is the only component that can generate intrinsic circular polarization, internal Faraday rotation, or contributes to the net magnetic flux of a jet.

\subsection{The amount of Unidirectional Magnetic Field}

We can constrain the amount of unidirectional magnetic field in a jet from flux conservation arguments. If the energy in a jet is carried by Poynting flux, then (e.g. Celotti et al. 1998):
\begin{equation}
B \sim 2 \times 10^{4} \left(\frac{L_{jet}}{10^{46} \,{\rm erg.s^{-1}}} \right)^{\frac{1}{2}} \left(\frac{\phi}{0.1 \,{\rm rad}} \right)^{-1} \left(\frac{R}{10^{14} \,{\rm cm}}\right)^{-1}\left(\frac{ \Gamma}{10}\right)^{-1}  {\rm Gauss}
\end{equation}
where $L_{jet}$ is the jet luminosity, $\phi$ is the opening angle, $R$ is the distance from the apex, and $\Gamma$ is the jet Lorentz factor.

A similar value is suggested by the ``Eddington field,'' which, attached to a maximally rotating black hole, radiates at the Eddington limit (Begelman, Blandford and Rees, 1984):
\begin{equation}
B \sim \left( \frac{8 \pi c^4 m_p}{\sigma_T G M} \right)^{\frac{1}{2}}  \sim 6 \times 10^4 \left( \frac{M}{10^8 M_{\odot}} \right)^{\frac{1}{2}} {\rm Gauss}
\end{equation}
If, close to the black hole, the field is mainly unidirectional, then the magnetic flux would be $\Phi \sim \pi R_{s}^2 B = 6 \times 10^{31} (M/10^{8} M_{\odot})^{3/2}$ Gauss cm$^{2}$. This is presumably a conserved quantity, and probably also an upper limit. We shall denote the uniform and unidirectional component of the magnetic field by $B_{uu}$. The double subscript emphasizes that this is not the same as what is frequently denoted by $B_u$, an ordered magnetic field component in which the field lines are parallel to each other, but do not necessarily point in the same direction. $B_{uu}$ and $B_{u}$ are indistinguishable in Stokes I and Q. The distinction is crucial in Stokes U (through Faraday rotation) and V.

If magnetic flux is conserved, then $B_{uu} \propto$ (jet radius)$^{-2}$. This falls off very quickly with distance from the jet apex, and the dominant field in most of the jet is a tangled component, $B_r$, consisting of disordered field and sheared loops that carry no net magnetic flux (Begelman, Blandford and Rees 1984). Both observationally, and in most theoretical models, $B_r \propto$ (jet radius)$^{-1}$. The ratio between these, $f_B$, is a useful quantity. We have $f_B = B_{uu}/B_r \propto $ (jet radius)$^{-1} \propto T_B^{\frac{1}{2}}$. The connection to brightness temperature follows approximately if we are comparing sources or features of similar flux density.

As an illustration, imagine that $f_B \approx 1$ at the base of the jet of a $ 3\times 10^8 M_{\odot}$ black hole, where (jet radius)$ \sim R_s \sim 10^{14}$  cm. Then at jet radii of $10^{16}$ and $10^{17}$ cm, we have $f_B$ in the range $10^{-2}$ to $10^{-3}$. For jet opening angles of a few degrees, this corresponds to distances from the base of a jet of a parsec or so, and this is roughly the linear resolution of the VLBA at short centimeter wavelengths for sources at intermediate redshifts.

This is a long way of saying that for the sources in which we have detected circular polarization close to or in the core component, we expect $f_B$ to be small, probably less than $1\%$. If $f_B$ were much larger than this, then conservation of magnetic flux would demand unreasonably strong magnetic fields at the base of the jet. 

The dependence of $f_B$ on jet radius and brightness temperature is very suggestive. If CP is produced either by the intrinsic mechanism, or by Faraday rotation driven Faraday conversion (see section 4), then we require a significant component of $B_{uu}$. Other things being equal, this will be most prominent in the most compact, highest brightness temperature sources. It may explain why so far we have only detected CP on the core, or emerging from the core (Homan and Wardle 1999), and why characteristic levels of CP are larger for the most active OVV's than for more typical AGN (Wardle and Homan 2001). It may also explain why the highest level of CP is found in an intra-day variable source, PKS 1519-273, for which $T_B \sim 5 \times 10^{14}K$ (Macquart et al 2000).

\section{Intrinsic Circular Polarization}

The intrinsic circular polarization of synchrotron radiation was worked out by Legg and Westfold (1968). In a uniform, unidirectional magnetic field, the fractional CP is of order $1/\gamma = (\nu/\nu_B)^{-1/2}$, where $\gamma$ is the Lorentz factor of the electrons radiating at frequency $\nu$, and $\nu_B = 2.8 B$ MHz is the electron gyrofrequency. More precisely (Jones and O'Dell 1977):
\begin{equation}
m_C = \frac{V}{I} = \epsilon_{\alpha}^{V} (\nu_{B\perp}/\nu)^{1/2}\cot\theta^{\prime}
\end{equation}
where $\nu_{B\perp} = eB\sin\theta^{\prime}/2\pi mc$, and $\theta^{\prime} = 90-\epsilon$ is the angle between the magnetic field and the line of sight in the frame of the radiating fluid. It is related to the angle between the jet and the line of sight, $\theta$, by the aberration formula $\sin\theta^{\prime} = D \sin\theta$. $D$ is the relativistic Doppler factor $D = [\Gamma(1 - \beta_{jet}\cos\theta)]^{-1}$, where the jet velocity is $\beta_{jet}c$ and $\Gamma = (1-\beta_{jet}^{2})^{-1/2}$ is its Lorentz factor. If the jet is oriented at the angle that maximizes superluminal motion, then $\theta^{\prime} = 90^{\circ}$ and $\epsilon = 0^{\circ}$. 

In equation (7), $\epsilon_{\alpha}^{V}$ is a dimensionless function of spectral index tabulated by Jones and O'Dell. A value of 1.8 covers the spectral range $\alpha = 0.5$ to $\alpha = 1.0$ to $10\%$ accuracy. Jones and O'Dell also include the effect of an anisotropic pitch angle distribution. We ignore that here because, as they point out, plasma instabilities do not permit an anisotropic distribution to be maintained. It should be noted that the divergence at small $\theta$ is rather misleading. Both V and I go to zero at small $\theta^{\prime}$ because of the $B_{\perp}$ dependencies in the V and I emissivities. In practice the I emission will then be dominated by any disordered field present, and $m_C$ will also go to zero. It should also be noted that the above expression assumes that the radiating particles are all electrons or all positrons, but not a mixture.

\subsection{Intrinsic CP in a partially ordered Field}
To illustrate these points, we consider a more realistic magnetic field structure consisting of an isotropic tangled field, $B_r$, plus a uniform field component, $B_u$, aligned with the jet. The unidirectional part of $B_u$ is $B_{uu} = f_B B_u$. The charge asymmetry is defined by $f_C = (n^- - n^+)/(n^- + n^+)$, where $n^-$ and $ n^+$ are the electron and positron densities, respectively. We combine these two factors into a single term, $\Lambda = f_B f_C$. The $\Lambda$ term should always be included, or at least considered, in any discussion of circular polarization or Faraday rotation. (It would be very useful if there were an observation that was sensitive to either $f_B$ or $f_C$ alone, rather than their product. We cannot think of one.) The fractional intrinsic circular polarization is now given by:
\begin{equation}
m_{C}^{int} = \epsilon_{\alpha}^{V} (\nu_{B\perp}/\nu)^{1/2}\frac{B_u \sin\epsilon}{<B_{\perp}^2>^{1/2}}\;\Lambda
\end{equation}
We have added a superscript to indicate that this is intrinsic CP rather than Faraday conversion, and reverted to $\epsilon = 90 -\theta^{\prime}$. Note that the divergence at small $\theta^{\prime}$, $\epsilon$ close to $90^{\circ}$ has disappeared. 

We can estimate the value of $B_{u}/<B_{\perp}^2>^{1/2}$ from the linear polarization. Assuming that $\xi$ ($=B_{u}/B_{r}$) is small, consistent with the modest fractional linear polarization typically observed, we can write $m_L = m_{0}\frac{3}{2}\xi^2\cos^{2}\epsilon$, and $<B_{\perp}^{2}> = \frac{2}{3}B_{r}^{2}$, so
\begin{equation}
m_{C}^{int} = \epsilon_{\alpha}^{V} (\nu_{B\perp}/\nu)^{1/2}\sqrt{\frac{m_{L}}{m_{0}}}\;\Lambda\tan\epsilon
\end{equation}
The divergence has not reappeared, because when $\epsilon = 90$, $m_{L} = 0$. Note that the connection to linear polarization is for this particular field model, a fine-grained tangled field plus a uniform component. For other models, it may be slightly different.

We see there are several factors that conspire to make intrinsic CP much smaller in real sources than is suggested by the equatiom (7). Quite apart from the possibility of a pair plasma, these include the degree of disorder in the magnetic field (which can be estimated by the linear polarization), and the fraction of the ordered component that is also unidirectional (which might be constrained by flux conservation arguments). Appreciable optical depth also suppresses both CP and LP. We do not treat that here because analytic expressions are very cumbersome, and the reader is referred to Jones and O'Dell (1977).

\subsection{The Connection with Faraday Rotation}

Faraday rotation also depends on the line of sight component of the magnetic field, its direction, and the sign of the charge on the particles. In the absence of thermal material in the source (Celotti et al 1998) internal Faraday rotation in a synchrotron source is caused by the lowest energy relativistic electrons. An electron of Lorentz factor $\gamma$ is less efficient at causing Faraday rotation than a thermal electron by a factor $\sim \gamma^{-2}$, because it is, in effect, heavier.

The Faraday depth, $\tau_F$, of a single cell of optical depth $\tau$, containing a unidirectional magnetic field is given by Jones and O'Dell (1977) as $\tau_{F} = \xi_{V}^{*} \tau$, where 
\begin{equation}
\xi_{V}^{*} = \xi^{*\,V}_{\alpha} (\nu/\nu_{B\perp})^{\alpha+\frac{1}{2}}\;\frac{\ln\gamma_{i}}{\gamma_{i}^{-(2\alpha+2)}} \cot\theta^{\prime}
\end{equation}
Here, $\gamma_i$ is the low energy cut off in the electron energy distribution, and $\xi^{*\,V}_{\alpha}$ is another dimensionless function of spectral index tabulated by Jones and O'Dell. It is equal to $1.27$ for $\alpha = 0.5$, and $0.75$ for $\alpha = 1.0$. For the magnetic field model used in the previous section, and averaging over many cells along the line of sight, we have
\begin{equation}
\tau_F = \xi^{*\,V}_{\alpha} (\nu/\nu_{B\perp})^{\alpha+\frac{1}{2}}\;\frac{\ln\gamma_{i}}{\gamma_{i}^{-(2\alpha+2)}}\; \frac{B_u \sin\epsilon}{<B_{\perp}^2>}^{1/2}\Lambda \tau
\end{equation}

We see that $\tau_F$ and $m_{C}^{int}$ depend in the same way on the field statistics and on $\Lambda$. It can be useful to combine the two expressions, to eliminate these poorly known quantities. Thus, if at some wavelength we measure a Faraday depth $\tau_F$ (which is a rotation in electric vector position angle of $\tau_{F}/4$ radians) then the expected intrinsic CP is
\begin{equation}
m_{C}^{int} = \frac{\tau_F}{\tau}\frac{\epsilon_{\alpha}^{V}}{\xi^{*\,V}_{\alpha}}(\nu/\nu_{B\perp})^{-\alpha-1}\;\frac{\gamma_{i}^{-(2\alpha+2)}}{\ln\gamma_{i}}
\end{equation}
This expression may be useful for constraining the value of $\gamma_i$. Note that $\tau_F$ refers to internal Faraday rotation. The observed Faraday depth includes any rotation external to the jet, and it may be difficult to disentangle these.

\section{Circular Polarization from Faraday Conversion}

This is a less familiar way of generating circular polarization in a synchrotron source, but it is likely to be more important than the intrinsic mechanism in many cases. This is especially true in a source dominated by a pair plasma (e.g. Wardle et al. 1998). 

Faraday conversion is a birefringence effect that converts LP to CP. In a weakly anisotropic thermal plasma the normal modes for electromagnetic waves are circular (Stokes V and -V) to a high degree of accuracy. Left and right circularly polarized waves propagate at different speeds, leading to the familiar Faraday {\em rotation} of linearly polarized radiation (Stokes Q is converted to Stokes U, and vice versa). In a relativistic plasma, the normal modes are noticeably elliptical. The circular component provides Faraday rotation, as before (see section 3.1). The normal modes for the linear component are Stokes Q and -Q, i.e. transverse and parallel to the magnetic field. Hence linearly polarized radiation at any other position angle (non-zero Stokes U) can be resolved into components transverse and parallel to the field, which propagate at different speeds. On emerging from the source, there is a phase difference between the two linearly polarized components which corresponds to circular polarization. This is exactly like applying the same sine wave to the horizontal and vertical plates of an oscilloscope, and introducing a phase difference between them. With increasing phase shift, Stokes U is progressively converted to V (and then to -U, -V and back to U etc). An important point is that this effect does not depend on the sign of the electron charge, or on the direction of the magnetic field. In contrast to the intrinsic mechanism, conversion occurs equally readily in an electron-proton plasma, a pair plasma, a unidirectional magnetic field, or a sheared field with zero net flux.

Synchrotron radiation in a uniform field only produces Stokes Q (and -Q at high optical depth). These are normal modes and there is no conversion without additional affects. One way to convert Q to U is through internal Faraday rotation, and this is the case treated in detail by Jones and O'Dell (1977). In a source with a uniform, unidirectional magnetic field, the fractional CP due to conversion is
\begin{equation}
m_{C}^{con} = \frac{1}{6}\epsilon_{Q} \tau_{c} \tau_{F}
\end{equation}
where $\epsilon_{Q} = (3\alpha +3)/(3\alpha + 5)$ is the fractional linear polarization (and is the same as $m_0$ in section 2), $\tau_{F} = \xi_{V}^{*}\tau$ is the Faraday depth (section 3.1), and $\tau_{c} = \xi_{Q}^{*}\tau $ is the ``conversion depth.'' The coefficient $\xi_{Q}^{*}$ is
\begin{equation}
\xi_{Q}^{*} = \xi^{*\,Q}_{\alpha} [(\nu/\nu_{i})^{\alpha-\frac{1}{2}}-1]/(\alpha-\frac{1}{2})
\end{equation}
where $\nu_{i} = \gamma_{i}^{2} \nu_{B\perp}$, and $\xi_{Q}^{*} = \xi^{*\.Q}_{\alpha} \ln(\nu/\nu_{i})$ if $\alpha = \frac{1}{2}$. $\xi^{*\,Q}_{\alpha}$ is another dimensionless function of spectral index tabulated by Jones and O'Dell. It is equal to $0.48$ for $\alpha = 0.5$ and $0.30$ for $\alpha = 1.0$.

Equation (13) can be interpreted qualitatively as follows. $\epsilon_{Q}$ represents the amount of linear polarized radiation (Stokes Q). $\tau_{F}$ represents how much of that is converted to U. $\tau_{C}$ represents how much of that U is then converted to V. Since $\tau_{F} \propto \lambda^2$ and $\tau_{C} \propto \lambda^3$, the spectrum of CP produced in this way is extremely steep at short wavelengths. At longer wavelengths, opacity, and Faraday depolarization of the linear polarization, both serve to suppress the CP.

\subsection{Faraday Conversion in a Partially Ordered Field}
We consider the same field model as in section 3.1: a fine grained tangled field of amplitude $B_r$, plus a uniform field $B_u$, Now $\epsilon_{Q}$ should clearly be replaced by $m_L$. Both $\tau_{c}$ and $\tau_{F}$ are also reduced by the field disorder, but we consider $\tau_{F}$ to be an observed quantity (despite the caveat at the end of section 3.2), and will keep it as such in our equations. The conversion depth, $\tau_{c}$, depends on the component of $B_u$ perpendicular to the line of sight, i.e. $\tau_c \propto  B_u^2\cos^2\epsilon/<B_{\perp}^2>\; \propto m_L/m_0$, and we can write
\begin{equation}
m_{C}^{con} = \frac{1}{6}\frac{m_L^2}{m_0} \;\tau_{F}^{obs} \;\xi_{Q}^{*} \;\tau
\end{equation}
For general optical and Faraday depth, see Ruszkowski and Begelman (2000).

\subsection{CP from Stochastic Faraday Conversion}
Faraday conversion operates on Stokes U. It was assumed above (and in most other treatments) that Q is converted to U by Faraday rotation, requiring a non-zero value of $\Lambda$, which precludes a pure pair plasma. There is another, and in some ways simpler way of producing U, and that is if the magnetic field fluctuates in orientation from cell to cell along the line of sight. Then what is radiated as ``Q'' in one cell is seen to have a component of ``U'' by the next cell, and conversion can take place. This can be quite efficient, and is akin to the way CP was produced in Jones' (1988) numerical simulations. 

This is a stochastic process, because in an isotropic tangled field, the resulting CP is equally likely to be positive or negative. It has a zero mean, but a non-zero root mean square value. Because of this, it cannot account for CP with long term sign stability, but it may be relevant to the variable part of the CP signal. It is most efficient in a coarse grained magnetic field, where the number of cells along a line of sight is not too large. 

The details of the calculation will be given elsewhere, but we find that for an isotropic field in which there are N cells along a line of sight,
\begin{equation}
m_C^{rms} \approx \frac{2}{5N} \;\epsilon_{Q}\; \xi_{Q}^{*}\; \tau 
\end{equation}
If there are $M$ independent lines of sight within a beam, then $m_C^{rms}$ is reduced by a further factor of $\sqrt{M}$.

Interestingly, the amount of CP is enhanced in a weak shock or compression, where the fluctuations in magnetic field direction from cell to cell are reduced (a $45^{\circ}$ change in azimuthal angle gives the largest amount of conversion). A strong shock reduces the CP because from the side ($\epsilon = 0$), the field appears completely ordered. If the compression is such that unit length is compressed to length $k$, and writing $x = (1-k^2)\cos^{2}\epsilon$, we find

\begin{equation}
m_C^{rms} \approx \frac{2}{5N} \;\epsilon_{Q}\; \xi_{Q}^{*} \;\tau (1-x)\;\left(1 + \frac{5Nx^2}{36}\right)^{1/2}
\end{equation}
In effect, the increasing correlation in the magnetic field turns this from a $1/N$ process to a $1/\sqrt{N}$ process.

Finally, we note that a jet with a helical magnetic field can be special case of Faraday conversion without Faraday rotation, loosely corresponding to $N=2$. If the field has a pitch angle $\phi$, then Stokes Q from the back half of the jet appears to the front half of the jet as $U = Q\sin2\phi$, and conversion can be extremely efficient. For a line of sight through the center of the jet, and at $\epsilon = 0$,
\begin{equation}
m_C \approx \epsilon_Q\; \xi_{Q}^{*} \;\frac{\tau}{4} \;\sin2\phi
\end{equation}
which can be large. It may be that significant circular polarization is a useful signature of a helical magnetic field.

\section{Circular Polarization from a Blandford-K\"{o}nigl Inhomogeneous Jet, and its Spectrum}

So far we have discussed CP from homogeneous sources. Real jets are presumably inhomogeneous, with quantities such as electron density and magnetic field strength varying with distance from the base of the jet. The simplest model of an inhomogeneous jet is due to  Blandford and K\"{o}nigl (1979), and we shall use this to illustrate some of the consequences for the production of CP. (To see some added complications of real jets, see Homan and Wardle, these proceedings)

The model describes a jet that is essentially self-similar over a wide range of wavelengths. This leads to a flat spectrum for Stokes I, and as we show below, a flat or even inverted spectrum for Stokes V. In the original version of the model, the jet shape is a cone, the magnetic field varies with distance, $r$, from the apex as $B \propto 1/r$, and the electron density varies as $ r^{-2}$. The jet is assumed to be in equipartition, which requires that it is also isothermal. The relativistic electron distribution is taken to be $N(\gamma) = K\gamma^{-2}$, for $\gamma_i< \gamma < \gamma_j$, so $K \propto r^{-2}$, but $\gamma_i$ and $\gamma_j$ are constant down the jet.

The optical depth varies as $\tau \propto \lambda^{3}/r^{3}$, so the base of the jet is always opaque, and the position, $r_1$, of the $\tau = 1$ surface varies as $r_{1} \propto \lambda$. The brightest part of the jet is close to the $\tau = 1$ surface, so at different wavelengths we are looking at different parts of the jet. This leads to a variety of somewhat unexpected frequency dependent effects, especially in the polarization.

One of these pointed out by Blandford and K\"{o}nigl was an apparent lack of internal Faraday rotation, despite a finite Faraday depth. This comes about as follows. The rotation measure at any point in the jet $\propto KBr \propto r^{-2}r^{-1}r \propto r^{-2}$. In the vicinity of $r_1$, the polarization position angle is rotated by an amount $\propto r_1^{-2} \lambda^2$, which is a constant. Thus the rotation angle in an unresolved jet is can be finite yet constant with wavelength. This may account for some VLBI cores that appear to have polarization position angles that are neither parallel nor perpendicular to the jet direction.

Electrons of energy $\gamma mc^2$ in a magnetic field B radiate most of their energy at a wavelength $\lambda \propto (\gamma^2 B)^{-1}$. It follows that the electrons at $r_1$ dominating the radiation at $\lambda$ have Lorentz factors given by $\gamma^2 \propto 1/B\lambda \propto r/\lambda$ which is constant. This means we are looking essentially at the same energy electrons at every wavelength. It follows that any intrinsic circular polarization from the jet has a flat spectrum, since $m_{C}^{int} \sim 1/\gamma$ which is constant, rather than the $\lambda^{\frac{1}{2}}$ dependence expected for a homogeneous source. 

The expected wavelength dependence of circular polarization due to Faraday conversion is also flat. If the Faraday depth, $\tau_F$, and conversion depth, $\tau_c$, are both small, then $m_{C}^{con} \propto \tau_F \tau_c$. At constant optical depth ($\tau \sim 1$), the Faraday and conversion depths are functions only of $\nu/\nu_{i} = (\gamma/\gamma_{i})^2 $, which we have just shown is constant in the Blandford-K\"{o}nigl model. 

Strong frequency dependencies in $\tau_F$ and $\tau_c$ and hence in $m_C$ immediately reappear if the jets are not isothermal. For example, if $\gamma_{i}$ decreases with distance down the jet, due to adiabatic losses, then CP generated by conversion regains a steep spectrum (with a slope that depends on the details of the jet model). If, on the other hand, the jet is isothermal, but the unidirectional part of the magnetic field (which provides the necessary internal Faraday rotation) decays as $r^{-2}$ (section 2.1, and Ruszkowski and Begelman 2002), then both $\tau_F$ and $m_C^{con}$ will vary as $\lambda^{-1}$, i.e. the CP will have an inverted spectrum. CP from the intrinsic mechanism is proportional to $B_{uu}/<B_{\perp}>$, and will also have an inverted spectrum, with $m_{C}^{int} \propto \lambda^{-1}$. These considerations suggest, rather counter-intuitively, that one might profitably search for circular polarization at the {\em highest frequencies}, and as close as possible to the base of the jets.

\end{article}

\end{document}